\documentclass[final,3p,times,twocolumn]{elsarticle}
\usepackage{amssymb}
\usepackage{epsfig}
\usepackage{graphicx}
\usepackage{dcolumn}
\usepackage{bm}
\usepackage{tabularx}
\usepackage{hyperref}
\usepackage{chngpage}
\usepackage{xcolor}
\usepackage{multirow}
\usepackage{amsmath}
\usepackage{subfigure}
\usepackage{stfloats}
\usepackage{float}

\journal{Advances Materials}

\begin{document}

\begin{frontmatter}

\title{Phase stability, chemical bonding and mechanical properties of titanium nitrides: A first-principles study}

\author[label1]{Shuyin Yu}
\author[label1]{Qingfeng Zeng\corref{cor1}}
\author[label2,label3,label4]{Artem R. Oganov}
\author[label5]{Gilles Frapper}
\author[label1]{Litong Zhang}
\address[label1]{Science and Technology on Thermostructural Composite Materials Laboratory, School of Materials Science and Engineering, Northwestern Polytechnical University, Xi'an, Shaanxi 710072, PR China}
\address[label2]{Department of Geosciences, Center for Materials by Design, and Institute for Advanced Computational Science, State University of New York, Stony Brook, NY 11794-2100, USA}
\address[label3]{Moscow Institute of Physics and Technology, Dolgoprudny, Moscow Region 141700, Russia}
\address[label4]{School of Materials Science and Engineering, Northwestern Polytechnical University, Xi'an, Shaanxi 710072, PR China}
\address[label5]{IC2MP UMR 7285, Universit\'{e} de Poitiers - CNRS, 4, rue Michel Brunet - TSA 51106 - 86073 Poitiers Cedex 9, France}
\cortext[cor1]{The author to whom the correspondence should be addressed to: Qingfeng Zeng, Email: qfzeng@nwpu.edu.cn}

\begin{abstract}
We have performed first-principles evolutionary searches for all stable titanium nitrides and have found, in addition to the well-known rocksalt-type TiN, new ground states Ti$_3$N$_2$, Ti$_4$N$_3$, Ti$_6$N$_5$ at atmospheric pressure, and Ti$_2$N and TiN$_2$ at higher pressures. The latest nitrogen-rich structure presents encapsulated N$_2$ dumbbells with a N-N distance of 1.348 \AA{} at 60 GPa and TiN$_2$ is predicted to be mechanically stable (quenchable). Our calculations of the mechanical properties (bulk modulus, shear modulus, Young's modulus, Poisson's ratio, and hardness) are in excellent agreement with the available experimental data and show that the hardness of titanium nitrides increases with increasing nitrogen content. The hardness of titanium nitrides is enhanced by strengthening directional covalent bonds and disappearance of Ti-Ti metallic bonds. Among the predicted compounds, TiN$_2$ has the highest hardness of 27.2 GPa.
\end{abstract}


\end{frontmatter}

\section{Introduction}
The discovery of new ultra-incompressible, superhard materials with novel mechanical and electronic properties is of great fundamental interest and practical importance. Conventional superhard materials are often formed by light elements (B, C, N, O), such as diamond \cite{occelli2003properties}, $\gamma$-B$_{28}$ \cite{BArtem}, \textit{c}-BN \cite{zheng2005superhard}, B$_6$O \cite{he2002boron}, and BC$_5$ \cite{solozhenko2009ultimate}. An alternative is to combine these elements with transition metals \cite{holleck1986material,kaner2005designing}, e.g., OsB$_2$ \cite{ja043806y}, ReB$_2$ \cite{Chung20042007}, IrN$_2$ and OsN$_2$ \cite{young2006synthesis}. Here we explore the Ti-N system, which gives promise for materials with superior properties: chemical stability, thermal stability, oxidative resistance, good adhesion to substrate, high fracture toughness and high hardness, and these properties make them suitable for many applications, as a thin film or coating for high-speed cutting tools.

TiN was first separated by Story-Maskelyne \cite{bannister1941osbornite} from a meteorite, followed by numerous studies on the film growth \cite{dubois1994model,dubois1992infrared,sherman1990growth}, elasticity and mechanical properties \cite{liu2012structural,brik2012first,yang2009first}. The crystal structure of TiN has been intensively investigated, however its phase transition under pressure is still a controversy \cite{jing2005isostructural,ojha2007pressure,chauhan2008structural}. The other well-known compounds are Ti$_2$N and Ti$_3$N$_4$. $\varepsilon$-Ti$_2$N \cite{holmberg1962structural} (\textit{P4$_2$/mnm}) is the most stable structure at normal conditions, while the $\delta$' phase (\textit{I4$_1$/amd}) can only exist at high temperatures \cite{de1985order}. Ivashchenko et al. \cite{ivashchenko2012first} predicted phase transformation sequence $\varepsilon$-Ti$_2$N $\rightarrow$ Au$_2$Te-type $\rightarrow$ Al$_2$Cu-type with transitions occurred at 77.5 and 86.7 GPa, respectively. For the metastable Ti$_3$N$_4$, Kroll et al. \cite{kroll2004assessment} proposed that it adopts the CaTi$_2$O$_4$-type structure at normal conditions. With increasing pressure, it was proposed to transform first into the Zr$_3$N$_4$-type structure, and then into the cubic Th$_3$P$_4$-type structure.

Although titanium nitrides have been studied for decades, a detailed theoretical study of their phase equilibria, electronic and mechanical properties would help to reconcile controversies and possibly predict new technologically useful materials. This is exactly the purpose of the present paper. We hope this study will provide guidance for experimental groups aiming to synthesize those novel crystal structures.
\section{Computational Methodology}
To find all potential Ti$_x$N$_y$ structures in the binary Ti-N phase diagram, we performed an extensive computational search for the most stable ground-state structures with the lowest enthalpy at selected pressures (P=1 atm, 20 and 60 GPa), using the evolutionary algorithm approach as implemented in the USPEX code \cite{oganov2006crystal,oganov2011evolutionary} in its variable-composition mode \cite{ISI:000277362500013}, interfaced with the VASP density functional package \cite{kresse1996efficient}. This approach of crystal structure prediction is based on purely chemical composition with no experimental input required. In our searches, we allowed all possible compositions in the Ti-N system with structures containing up to 24 atoms in the unit cell. The first generation of structures was produced randomly, and the succeeding generations were obtained by applying heredity (50$\%$), atom transmutation (20$\%$), lattice mutation (15$\%$) operations and produced randomly (15$\%$).

The first-principles electronic structure calculations were carried out within the framework of density functional theory (DFT) with exchange-correlation functionals in the generalized gradient approximation (GGA) according to Perdew-Burke-Ernzerhof (PBE) parameterization \cite{perdew1996generalized}. The interactions between the ions and the electrons are described by the projector-augmented wave (PAW) method \cite{blochl1994projector} with a cutoff energy of 600 eV. Structures were optimized until the maximum residual force component was smaller than 1 meV/\AA{}. All the enthalpy calculations are well converged to an accuracy of $<$10$^{-6}$eV/atom. Brillouin zone was sampled by uniform $\Gamma$-centered Monkhorst-Pack meshes \cite{monkhorst1976special} with the resolution 2$\pi$$ \times $0.06 $ \AA{} $$^{-1}$. The most stable structures were studied further with increased accuracy. \textit{k}-points meshes of resolution better than 2$\pi$$ \times $0.03 $ \AA{} $$^{-1}$ were adopted.

Theoretical phonon spectra were calculated based on the supercell method using the PHONOPY package \cite{togo2008first} in order to probe the dynamic stability of the predicted Ti$_x$N$_y$ compounds at different pressures. The elastic constants were calculated from the stress-strain relations, and Voigt-Reuss-Hill (VRH) approximation \cite{hill1952elastic,voigt1928lehrbuch,reuss1929calculation} was employed to obtain the bulk modulus \textit{B}, shear modulus \textit{G}, Young's modulus \textit{E}, and Poisson's ratio \textit{$\nu$}. The theoretical Vickers hardness \textit{H$_v$} was estimated by using the Chen's model \cite{chen2011modeling}, according to the expression:
\begin{equation}\label{Equa1}
H_v = 2{({ \kappa ^2}G)^{0.585}} - 3
\end{equation}
where $\kappa$ is the Pugh ratio: $\kappa$=\textit{G}/\textit{B}. The directional dependence of the Young's modulus for crystals of different symmetries encountered in this study is given in Supplementary Materials.
\section{Results and Discussion}
\begin{figure}[H]
\centering
\begin{minipage}[b]{0.4\textwidth}
\centering
\includegraphics[width=2.5in]{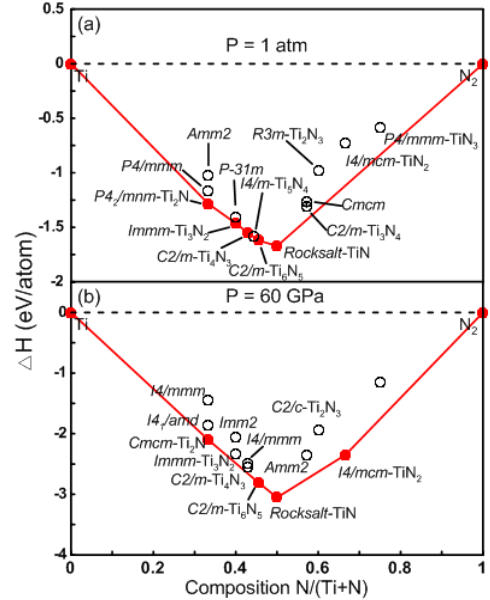}
\end{minipage}
\caption{(Color online). Convex hull diagrams for Ti-N system at ambient pressure (a) and at 60 GPa (b). Circles denote stable (solid circles) and metastable (open circles) structures. The hexagonal $\omega$-Ti \cite{vohra2001novel} and $\alpha$- N$_2$ \cite{bini2000high} are adopted as reference states. \label{Fig1}}
\end{figure}

Thermodynamics of titanium nitrides can be quantified by constructing the thermodynamic convex hull, which is a complete set of phases stable against transformation into any other phase and decomposition into any set of other phases. These thermodynamically stable phases can be synthesized in principle \cite{Ghosh20083202}. The convex hull curves were reconstructed after phonon calculations for the selected thermodynamically stable phases, which are shown in Figure \ref{Fig1}. Noticeably, our evolutionary simulations succeed in finding the well-known NaCl-type TiN and $\varepsilon$-Ti$_2$N observed by experiments at ambient conditions. In addition, we uncovered three new stable compounds, the orthorhombic Ti$_3$N$_2$, and monoclinic Ti$_4$N$_3$ and Ti$_6$N$_5$. At 60 GPa, $\varepsilon$-Ti$_2$N transforms into the orthorhombic \textit{Cmcm} phase, and a new stable nitrogen-rich compound is identified as TiN$_2$.

The detailed crystallographic data and enthalpies of formation are listed in Table 1s (see Supplementary Materials). Note that the enthalpies of formation without the ZPE correction are only $\sim$0.01 eV/atom different from the ZPE-corrected energies, and it will not change the relative phase stabilities without considering ZPE correction. Thus, for the computational intensity reason, the zero point energy is omitted for the systematic structural search.
\subsection{Rocksalt TiN and related subnitrides Ti$_{n+1}$N$_n$}
\begin{figure}
\centering
\begin{minipage}[b]{0.5\textwidth}
\centering
\includegraphics[width=3in]{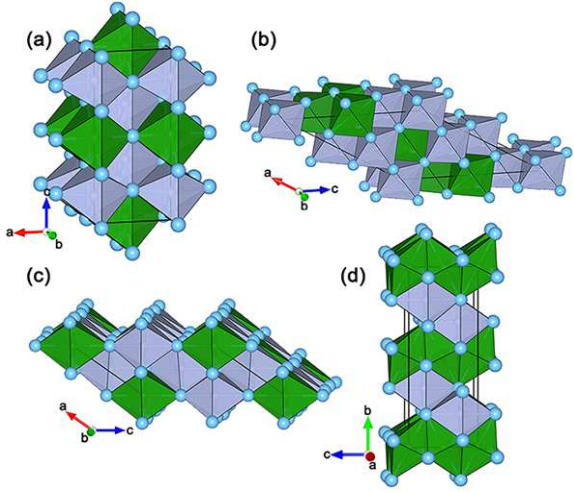}
\end{minipage}
\caption{(Color online). Crystal structures of subnitrides Ti-N compounds. (a) \textit{Immm}-Ti$_3$N$_2$ at 0 GPa, (b) \textit{C2/m}-Ti$_4$N$_3$ at 0 GPa, (c) \textit{C2/m}-Ti$_6$N$_5$ at 0 GPa, (d) \textit{Cmcm}-Ti$_2$N at 60 GPa. The Ti and N atoms are represented as big blue and small gray spheres, respectively, nitrogen vacancies are shown in the green octahedra.\label{Fig2}}
\end{figure}

While TiN has the ideal cubic rocksalt structure, Ti$_3$N$_2$, Ti$_4$N$_3$ and Ti$_6$N$_5$ are versions of this structure with ordered N-vacancies (see Figure \ref{Fig2}) - in Ti$_3$N$_2$ one third of nitrogen sites are vacant, in Ti$_4$N$_3$ - one quarter, and in Ti$_6$N$_5$ - one sixth. Similar vacancy-ordered phases were earlier reported to be stable for hafnium and titanium carbides M$_{n+1}$C$_n$
(M=Hf and n=2, 5 \cite{PhysRevB.88.214107}; M=Ti and n=1, 2, 5 \cite{jiang2014pressure}). Generally, the decrease of vacancy concentration makes the structure denser, less compressible, harder, and more stable under pressure. Indeed, among defective rocksalt-type phases only Ti$_6$N$_5$ (with the lowest vacancy concentration) survives as a (barely) stable phase at 60 GPa. Ti$_3$N$_2$ has space group \textit{Immm}, while Ti$_4$N$_3$ and Ti$_6$N$_5$ belong to space group \textit{C2/m}. It is convenient to visualize these structures by N-centered octahedral (NTi$_6$), as shown in Figure \ref{Fig2}. In the Ti$_{n+1}$N$_n$ series, Ti$_5$N$_4$ is missing as a stable compound. Nevertheless, a structure (SG: \textit{I4/m}, No. 87) is very close to the convex hull curve but lies above it (15 meV/atom), i.e., Ti$_5$N$_4$ is a metastable phase.

$\varepsilon$-Ti$_2$N has the anti-rutile structure with space group \textit{P4$_2$/mnm}. At pressures above 20.8 GPa, tetragonal $\varepsilon$-Ti$_2$N transforms into an orthorhombic \textit{Cmcm} structure (see Figure \ref{Fig2}e). This structure contains a distorted close packing of Ti atoms, where one half of octahedral voids are occupied by nitrogen atoms, and edge-sharing NTi$_6$-octahedra form double slabs alternating with nitrogen-free layers. It is noteworthy that the enthalpy of \textit{Cmcm}-Ti$_2$N is lower by more than 0.24 eV/f.u. than that of the Au$_2$Te-type or Al$_2$Cu-type structures proposed by Ivashchenko et al. \cite{ivashchenko2012first}.
\begin{figure}[H]
\centering
\begin{minipage}[b]{0.5\textwidth}
\centering
\subfigure{
\label{Fig3a}
\includegraphics[width=2.15in]{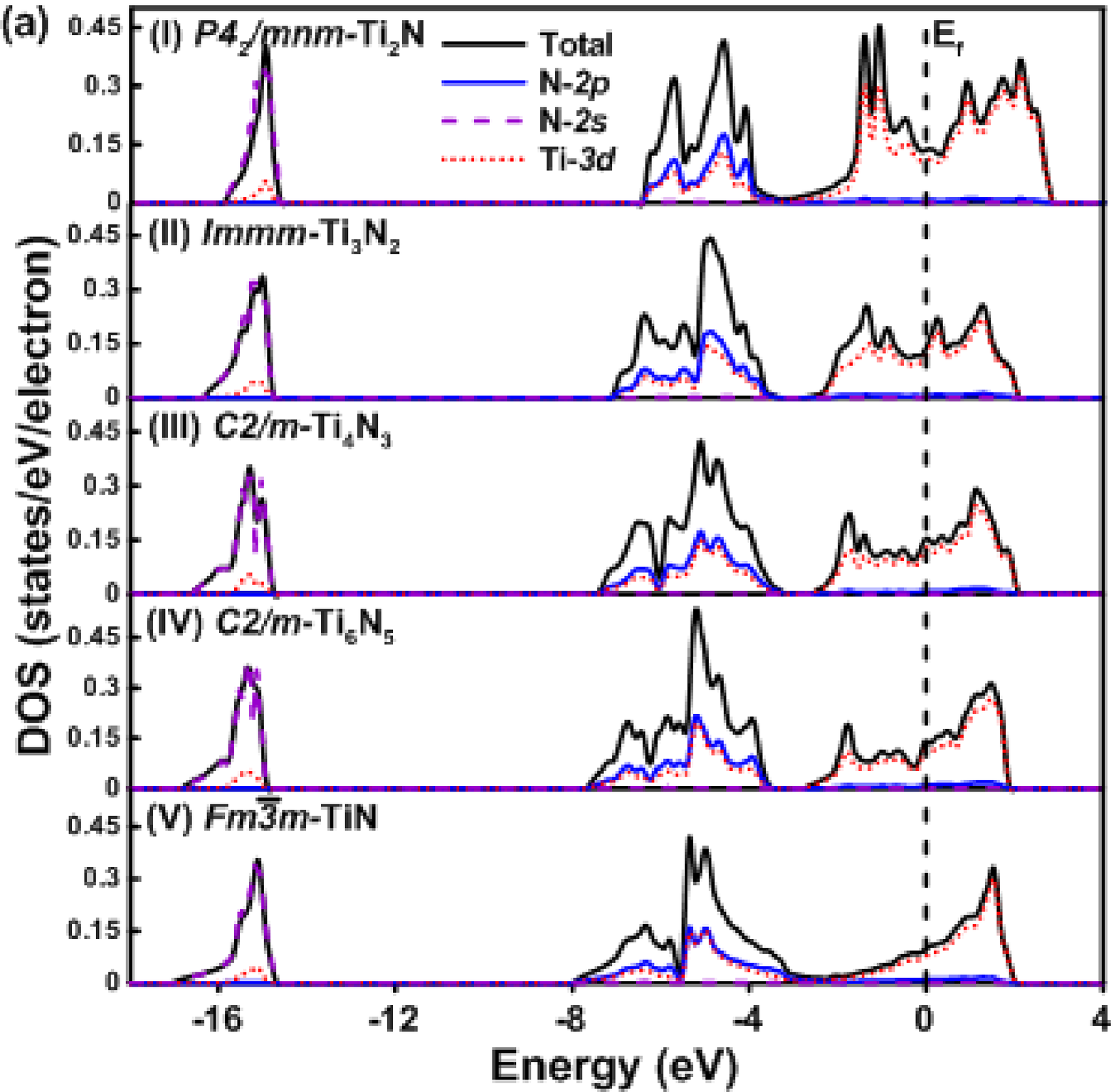}}
\subfigure{
\label{Fig3b}
\includegraphics[width=2.2in]{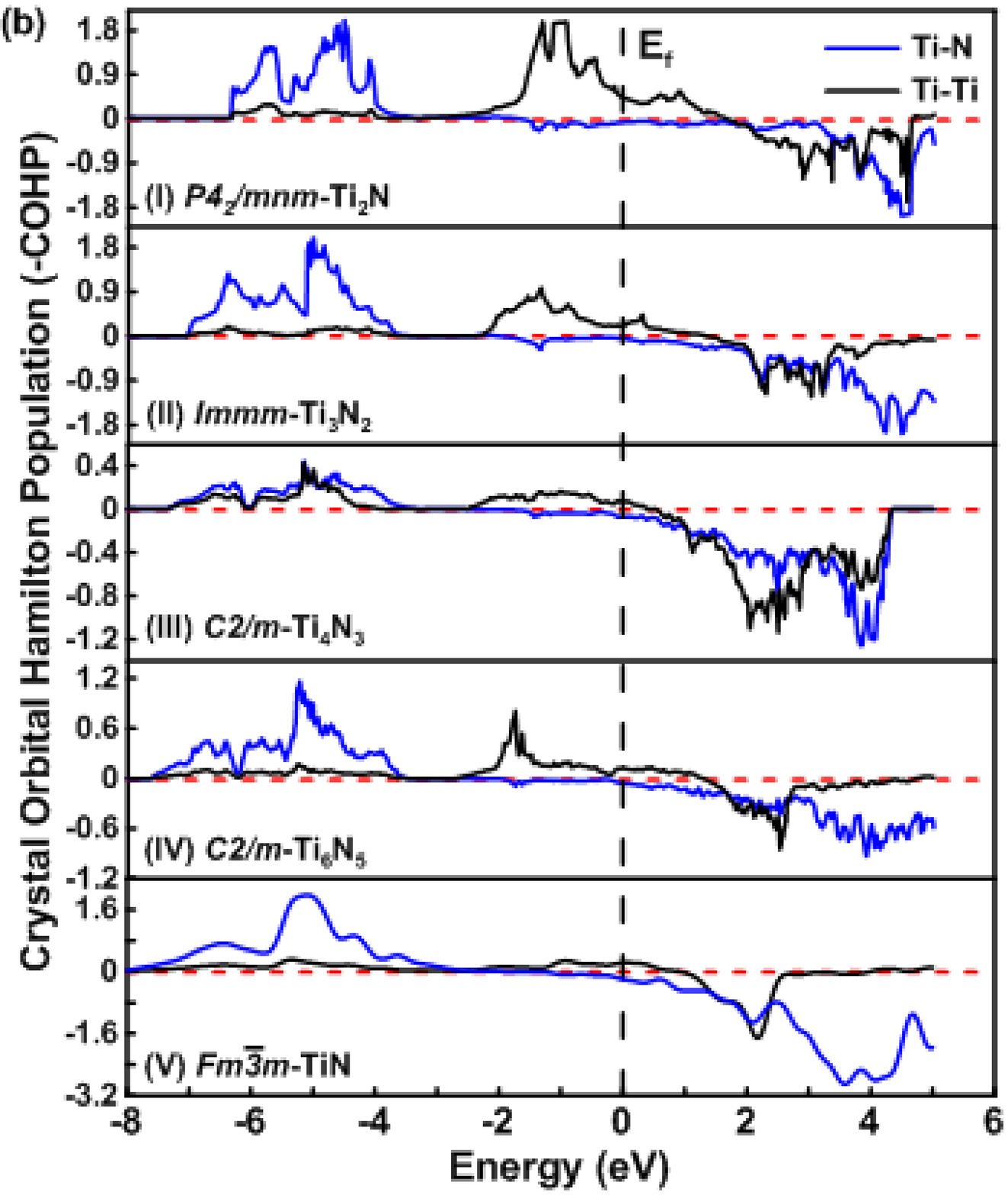}}
\end{minipage}
\caption{(Color online). (a) The total and partial density of states (DOS), (b) Crystal orbital Hamilton population (-COHP) for the Ti$_{n+1}$N$_n$ subnitrides and TiN at zero pressure. Vertical dashed line is the Fermi energy.}
\end{figure}

Having investigated the crystal structures and energetics of the various titanium subnitrides Ti$_{n+1}$N$_n$ and TiN, we now consider their electronic properties. The total and partial densities of states (DOS) of the understudied compounds are shown in Figure \ref{Fig3a}. The crystal orbital Hamilton population (-COHP) curves for all Ti-N compounds were calculated by using the LOBSTER package \cite{lobster01} and are displayed in Figure \ref{Fig3b}. These orbital-pair interactions can provide a quantitative measure of bond strengths. The positive and negative energy regions in the -COHP curves correspond to bonding and antibonding states, respectively. For Ti$_{n+1}$N$_n$ structures, the DOS are decomposed into three well separated regions energy: (1) a deeply lowest valence band, \textit{s}$_N$; (2) a hybridized Ti-\textit{3d}/N-\textit{2p} band, \textit{d}$_M$\textit{p}$_N$; (3) a partially filled higher lying Ti-\textit{3d} band, \textit{d}$_M$. The \textit{s}$_N$ band is dominated by the \textit{2s} orbitals of the nitrogen atoms. Therefore, the contribution of this band to the bonding is not so large. The following upper group of valence bands \textit{d}$_M$\textit{p}$_N$ is a result of strong hybridization from the \textit{3d} states of Ti atoms and the \textit{2p} states of N atoms. These peaks correspond to the Ti-\textit{3d}/N-\textit{2p} bonding orbitals contribution. Their antibonding counterparts appear well above the Fermi level.

When n goes from 1 to 5 in Ti$_{n+1}$N$_n$ series, e.g., the number of N-filled Ti$_6$ octahedra increases, the energy region for the Ti-\textit{3d}/N-\textit{2p} peaks is expanded. This energy dispersion is caused by more extensive mixing between Ti-\textit{3d} and N-\textit{2p} orbitals, and suggests the enhancement of the covalency of titanium-nitrogen bonding network. Let turn our discussion on the upper energy region. From both DOS and COHP analyses, one may see that the bottom of the \textit{d}$_M$ band is mainly dominated by the bonding orbitals of \textit{3d} Ti atoms and the corresponding DOS exhibits a feature of ``near free electron'' responsible for metallicity (see Figure \ref{Fig3a}). Vacant antibonding Ti-\textit{3d}/N-\textit{2p} and antibonding Ti-\textit{3d}/N-\textit{3d} states compose the bottom of the conduction band. These results confirm the mixed covalent Ti-N and metallic Ti-Ti bonding nature in titanium subnitrides and TiN compounds.
\begin{figure}[H]
\centering
\begin{minipage}[b]{0.5\textwidth}
\centering
\includegraphics[width=3in]{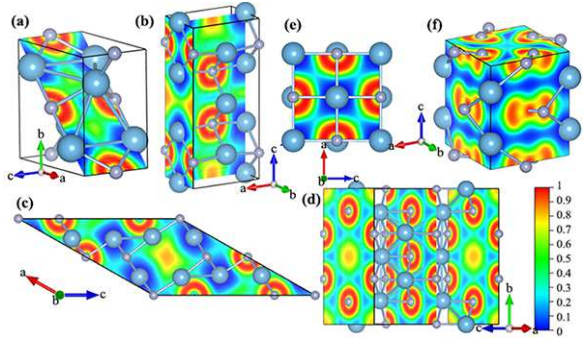}
\end{minipage}
\caption{(Color online). Calculated electron localization function (ELF) maps for (a) \textit{P4$_2$/mnm}-Ti$_2$N, (b) \textit{Immm}-Ti$_3$N$_2$, (c) \textit{C2/m}-Ti$_4$N$_3$, (d) \textit{C2/m}-Ti$_6$N$_5$, (e) \textit{Fm$\bar{3}$m}-TiN, and (f) \textit{I4/mcm}-TiN$_2$, blue spheres represent Ti atoms while gray spheres represent N atoms. \label{Fig4}}
\end{figure}

As N content increases, valence band broadens (\textit{d}$_M$\textit{p}$_N$ band), valence electron concentration increases, and covalent interactions become stronger. This is reflected in shortening of Ti-N bonds: the shortest bond lengths are 2.082, 2.071, 2.069 and 2.022 \AA{} for Ti$_2$N, Ti$_3$N$_2$, Ti$_4$N$_3$ and Ti$_6$N$_5$ at 0 GPa, respectively. In the same sequence the N(\textit{E$_F$}) gradually decreases, reaching the lowest value $\sim$0.098 states/eV/electron for TiN (see Figure \ref{Fig3a}). To further explore the bonding nature in each Ti$_{n+1}$N$_n$ and TiN structures, we examined the electron localization function (ELF) \cite{becke1990simple} of these Ti-N compounds (see Figure \ref{Fig4}). We find that there are relatively large ELF values between Ti and N atoms, indicating the partially covalent Ti-N interactions. ELF=0.5 can be identified between titanium atoms, suggesting that the metallic Ti-Ti bonding. All of these results are consistent with the discussion about the DOS and COHP for each structure of Ti$_x$N$_y$.
\subsection{TiN$_2$, a high-pressure structure with N$_2$ dumbbells}
\begin{figure}[H]
\centering
\begin{minipage}[b]{0.5\textwidth}
\centering
\includegraphics[width=3in]{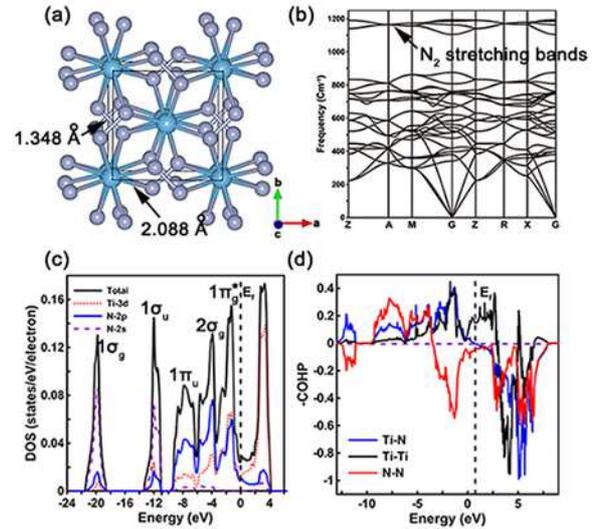}
\end{minipage}
\caption{(Color online). (a) Crystal structure of \textit{I4/mcm}-TiN$_2$, (b) phonon dispersion curve at 60 GPa, (c) projected and total DOS with symmetry labels of molecular N$_2$$^{4-}$ filled valence orbitals assigned to each DOS peak at 60 GPa, and (d) -COHP of \textit{I4/mcm}-TiN$_2$ at 60 GPa.\label{Fig5}}
\end{figure}
At 60 GPa, we uncovered a nitrogen-rich compound TiN$_2$ with the CuAl$_2$-type structure (SG: \textit{I4/mcm}, see Figure \ref{Fig5}a). The energy of this \textit{I4/mcm} structure is lower than that of the previously proposed thermodynamic ground state, the CaC$_2$-V-type structure predicted by Kulkarni et al. \cite{kulkarni2013structure} at 60 GPa. This new structure is stable against the decomposition into
the mixture of TiN and N$_2$ at pressures above 26.6 GPa. To examine the stability of the \textit{I4/mcm} structure, the phonon frequencies were calculated at atmospheric pressure and at 60 GPa. Imaginary frequency mode was not observed in the entire Brillouin zone (see Figure \ref{Fig5}b and Supplementary Materials), indicating that the predicted \textit{I4/mcm} structure is dynamical stable and, if synthesized, might be quenchable at atmospheric pressure.
\begin{table*}[bp]
\scriptsize
\begin{center}
\caption{\label{Tab1} The calculated elastic constants \textit{C$_{ij}$}, bulk modulus \textit{B} (GPa), shear modulus \textit{G} (GPa), Young's modulus \textit{E} (GPa), Poisson's ratio \textit{$\nu$}, \textit{$\kappa$}=\textit{G}/\textit{B} ratio, and Vickers hardness \textit{H$_v$} (GPa) of the Ti-N compounds at 0 GPa.}
\begin{tabular}{p{0.8cm}p{0.7cm}p{0.3cm}p{0.3cm}p{0.3cm}p{0.4cm}p{0.3cm}p{0.3cm}p{0.3cm}p{0.3cm}p{0.4cm}p{0.3cm}p{0.3cm}p{0.3cm}p{0.3cm}
p{0.3cm}p{0.3cm}p{0.3cm}p{0.3cm}p{0.3cm}p{0.4cm}}
\hline \hline
Phase & & \textit{C$_{11}$} & \textit{C$_{12}$} & \textit{C$_{13}$} & \textit{C$_{16}$} & \textit{C$_{22}$} & \textit{C$_{23}$} & \textit{C$_{26}$} & \textit{C$_{33}$}& \textit{C$_{36}$}& \textit{C$_{44}$}& \textit{C$_{45}$}& \textit{C$_{55}$}& \textit{C$_{66}$} &\textit{B}&\textit{G}&\textit{E}&\textit{$\nu$}&\textit{$\kappa$}& \textit{H$_v$}\\
\hline
Ti$_2$N                & GGA & 300& 214 & 115&     &    &    &    & 442&    & 157&    &    & 137&214& 112& 287& 0.277& 0.523& 11.87\\
(\textit{P4$_2$/mnm})  & LDA & 353& 235 & 138&     &    &    &    & 489&    & 184&    &    & 147&246& 130& 332& 0.275& 0.528& 13.38\\
                       & Ave & 327& 225 & 127&     &    &    &    & 467&    & 171&    &    & 142&230& 121& 310& 0.276& 0.526& 12.63\\
Ti$_3$N$_2$            & GGA & 458& 99  & 125&     & 400& 115&    & 332&    & 119&    & 145& 86 &206& 124& 309& 0.250& 0.602& 15.53\\
(\textit{Immm})        & LDA & 545& 113 & 140&     & 449& 140&    & 370&    & 127&    & 172& 92 &237& 138& 347& 0.256& 0.583& 16.00\\
                       & Ave & 502& 106 & 133&     & 425& 128&    & 351&    & 123&    & 159& 89 &222& 131& 328& 0.253& 0.593& 15.77\\
Ti$_4$N$_3$            & GGA & 368& 150 & 155& -17 & 393& 119& 3  & 421& -29& 140& 10 & 126& 114&225& 125& 317& 0.265& 0.556& 14.05\\
(\textit{C2/m})        & LDA & 412& 181 & 179& -16 & 444& 135& 8  & 484& -42& 167& 19 & 139& 130&258& 141& 357& 0.269& 0.547& 14.84\\
                       & Ave & 390& 166 & 167& -17 & 419& 127& 6  & 453& -36& 154& 15 & 133& 122&242& 133& 337& 0.267& 0.552& 14.45\\
Ti$_6$N$_5$            & GGA & 424& 140 & 145& -20 & 429& 146& 20 & 429& 1  & 147& 21 & 158& 171&238& 151& 373& 0.239& 0.634& 18.97\\
(\textit{C2/m})        & LDA & 484& 161 & 175& -33 & 490& 175& 33 & 481& 1  & 166& 34 & 187& 201&275& 170& 424& 0.243& 0.618& 20.07\\
                       & Ave & 454& 151 & 160& -27 & 460& 161& 27 & 455& 1  & 157& 28 & 173& 186&257& 161& 399& 0.241& 0.626& 19.52\\
TiN                    & GGA & 590& 145 &    &     &    &    &    &    &    & 169&    &    &    &294& 189& 466& 0.235& 0.643& 22.55\\
(\textit{Fm$\bar{3}$m})& LDA & 704& 157 &    &     &    &    &    &    &    & 183&    &    &    &339& 215& 533& 0.238& 0.634& 24.22\\
                       & Ave & 647& 151 &    &     &    &    &    &    &    & 176&    &    &    &317& 202& 500& 0.237& 0.639& 23.39\\
                       & GGA\cite{liu2012structural}  & 583& 129 &&&& & & & & 179&    &    &    &278&    &    &    & &24.19\cite{TiNhardness}\\
                       & LDA\cite{liu2012structural}  & 698& 139 &&&& & & & & 198&    &    &    &321&    &    &      &      &      \\
                       & Exp\cite{PhysRevB.53.3072}   & 625& 165 &&&& & & & & 163&    &    &    &320&    &    &&&23.00\cite{TiNhardnessexpt}\\
TiN$_2$                & GGA & 535& 279 & 71 &     &    &    &    & 653&    & 336&    &    & 148&284& 197& 481& 0.218& 0.693& 25.64\\
(\textit{I4/mcm})      & LDA & 631& 301 & 106&     &    &    &    & 729&    & 364&    &    & 185&335& 233& 567& 0.218& 0.696& 28.75\\
                       & Ave & 583& 290 & 89 &     &    &    &    & 691&    & 350&    &    & 167&310& 215& 524& 0.218& 0.695& 27.20\\
\hline
\end{tabular}
\end{center}
\end{table*}

The structure of TiN$_2$ contains TiN$_8$ face-sharing tetragonal antiprisms stacked along the \textit{c}-axis. Unlike in normal CuAl$_2$-type structures, here we have N$_2$-dumbbells encapsulated in cubic Ti$_8$ hexahedra. The N-N bond length is calculated as 1.348 \AA{} and 1.378 \AA{} at 60 GPa and 1 atm, respectively. Projected N-\textit{2p} DOS shows that antibonding 1$\pi$$_g$$^*$ levels are almost fully occupied at 60 GPa (see Figure \ref{Fig5}c). Consequently, for electron counting purposes, the dinitrogen units should be formally considered as N$_2$$^{4-}$, a pernitride unit isoelectronic to fluorine F$_2$ molecules. Its electronic ground state configuration is 1$\sigma$$_g$$^2$, 1$\sigma$$_u$$^2$, 1$\pi$$_u$$^4$, 2$\sigma$$_g$$^2$, 1$\pi$$_g$$^{*4}$ for 14 valence electrons and its bond order is one. Formally, this leaves the titanium atoms of TiN$_2$ in a \textit{d$^0$} configuration (Ti$^{4+}$). Note that a large variety of structures are known with the square antiprismatic geometry and a formally \textit{d$^0$} metal center such as TaF$_8$$^{3-}$ in Na$_3$TaF$_8$ \cite{burdett1978eight}.

Stabilizing Ti-Ti interactions in binary Ti-N structures are weakening when nitrogen content increases, e.g., when formal valence \textit{d} electrons \textit{d$^n$} decrease (Ti oxidation number increases). This behavior is structurally reflected by the elongation of the Ti-Ti distances upon the nitrogen content (shortest Ti-Ti bonds of 2.255 \AA{} and 2.517 \AA{} in Ti$_2$N and TiN$_2$ at 60 GPa, respectively). Note that the pernitride N$_2$$^{4-}$ units point directly towards the tetragonal faces of the eight-coordinated titanium atoms (Ti-N bonding length is 2.088 \AA{} at 60 GPa) and are perpendicular to each other in order to minimize the steric clashes between the nitrogen $\sigma$-lone pairs (Pauli repulsions). Based on these considerations, the covalent networks in TiN$_2$ with the CuAl$_2$-type structure are built on transition metal - main group elements and main-group-based directional  covalent bonds, a prerequisite for superhard materials.
\subsection{Mechanical properties of Ti-N stable structures}
The dynamical stability of the predicted Ti-N compounds has been confirmed by the absence of any imaginary phonon modes (see Supplementary materials). The mechanical stability was examined using the calculated elastic constants (see Table \ref{Tab1}), and we found that all these phases satisfy the Born-Huang stability criteria \cite{wu2007crystal}. Our calculated elastic constants are in good agreement with the available data \cite{liu2012structural,PhysRevB.53.3072,kim1992elastic,wang2010structural}. One can note the large \textit{C$_{33}$} values for Ti$_2$N, Ti$_4$N$_3$, TiN and especially TiN$_2$ (691 GPa), which indicate very low compressibility along the \textit{c}-axis. For TiN$_2$, we also find a very large elastic constant \textit{C$_{44}$} (350 GPa).

\begin{figure}
\centering
\begin{minipage}[b]{0.5\textwidth}
\centering
\includegraphics[width=2.2in]{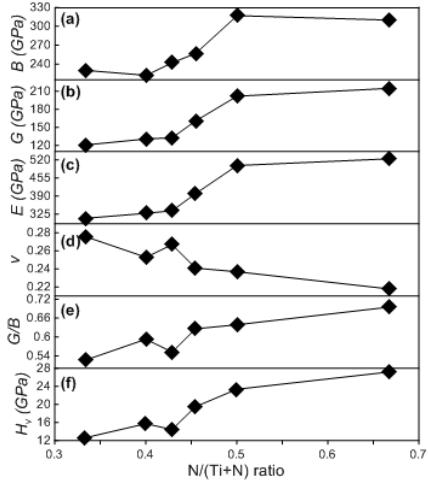}
\end{minipage}
\caption{(Color online). Calculated bulk modulus \textit{B}, shear modulus \textit{G}, Young's modulus \textit{E}, Poisson's ratio \textit{$\nu$}, \textit{G}/\textit{B} ratio, and Vickers hardness \textit{H$_v$} of titamium nitrides as a function of N content. The structures from left to right are $\varepsilon$-Ti$_2$N, \textit{Immm}-Ti$_3$N$_2$, \textit{C2/m}-Ti$_4$N$_3$, \textit{C2/m}-Ti$_6$N$_5$, \textit{Fm$\bar{3}$m}-TiN, and \textit{I4/mcm}-TiN$_2$.\label{Fig6}}
\end{figure}

The calculated bulk modulus \textit{B} and shear modulus \textit{G} are also listed in Table \ref{Tab1} and the trend of these mechanical properties as a function of N content is shown in Figure \ref{Fig6}. Our calculated results for the well-known rocksalt TiN and $\varepsilon$-Ti$_2$N are in good agreement with reported values \cite{PhysRevB.53.3072,liu2012structural,kim1992elastic,wang2010structural}, supporting the accuracy and reliability for other Ti-N compounds. TiN has the largest \textit{B} of 317 GPa, slightly larger than TiN$_2$ of 310 GPa. Both values are very high and comparable to WB$_4$ with a bulk modulus of 328 GPa \cite{WB4}. Bulk modulus is mainly related to valence electron density (VED) \cite{Gilman20061}, and indeed the VED of TiN$_2$ (0.478 \textit{e$^-$}/\AA{}$^3$) is very close to the WB$_4$ (0.485 \textit{e$^-$}/\AA{}$^3$). A minimum of \textit{B} appears in Ti$_3$N$_2$, because this is a vacancy-rich rocksalt-type phase, and due to the large concentration of vacancies it has unexpectedly low \textit{C$_{12}$} and \textit{C$_{22}$} values as listed in Table \ref{Tab1}. Shear modulus measures the resistance to shape change at constant volume and provides a much better correlation with hardness than bulk modulus, and TiN$_2$ has the largest shear modulus of 215 GPa.

\begin{figure}
\centering
\begin{minipage}[b]{0.5\textwidth}
\centering
\includegraphics[width=2.8in]{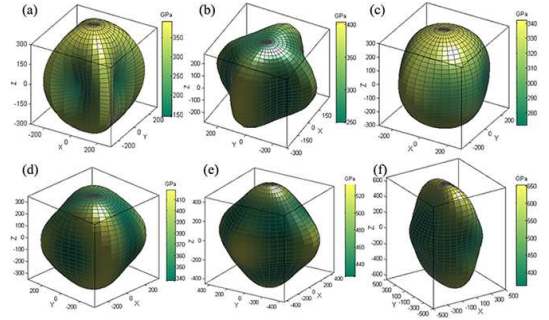}
\end{minipage}
\caption{(Color online). Directional dependence of Young's moduli (in GPa) for (a) \textit{P4$_2$/mnm}-Ti$_2$N, (b) \textit{Immm}-Ti$_3$N$_2$, (c) \textit{C2/m}-Ti$_4$N$_3$, (d) \textit{C2/m}-Ti$_6$N$_5$, (e) \textit{Fm$\bar{3}$m}-TiN, and (e) \textit{I4/mcm}-TiN$_2$. \label{Fig7}}
\end{figure}

Figure \ref{Fig7} shows the directional dependence of Young's moduli of the considered Ti-N compounds, and one can clearly see anisotropy from the deviations of its shape from sphere. The Young's modulus along the \textit{c}-axis direction is higher than that along the other directions for Ti$_2$N, Ti$_4$N$_3$, and TiN$_2$, which agrees well with our above analysis on elastic constants. The Young's modulus is more isotropic in Ti$_4$N$_3$, while TiN$_2$ shows more anisotropy than other Ti-N compounds, because of its high \textit{$C_{44}$}.

Poisson's ratio \textit{$\nu$} and \textit{G/B} ratio \cite{pugh1954xcii} are indicative of the degree of directionality of the covalent bonding. A typical value for \textit{$\nu$} is $\sim$0.2 for strong directional covalent materials and $\sim$0.4 for good metals. A low Poisson's ratio results from directional bonds, which increase the shear modulus and limit the motion of dislocations, thereby increasing a material's hardness. From Table \ref{Tab1}, we find that \textit{$\nu$} drops down from 0.276 in Ti$_2$N to 0.218 in TiN$_2$ with increasing N content, except Ti$_4$N$_3$, which shows a local maximum. For Ti$_4$N$_3$, we can find that there are large Ti-Ti metallic bonding in contrast with other Ti-N compounds (see Figure \ref{Fig3b}). The large Ti-Ti metallic bonds impair the covalency of this phase, which lead to a local minimum of hardness (see Figure \ref{Fig6}f). The Vickers hardness was estimated by Chen's empirical method and listed in Table \ref{Tab1}. Among the studied Ti-N compounds, TiN$_2$ has the largest hardness of 27.2 GPa. Note that this is an orientationally averaged value (anisotropy is not considered in Equation \ref{Equa1}). It is no wonder that increasing nitrogen content changes bonding to more directional, the more N is added, the more Ti-N bonds are formed and the less Ti-Ti bonding remains. Together with the analyses on electronic properties and chemical bonding in those stable Ti-N compounds, We can conclude that with increasing N content, the enhancement of directional covalent interactions and decline of metallicity lead to the increase of the hardness.
\section{Conclusions}
In summary, we have extensively explored the stable structures and possible stoichiometries in the Ti-N system by first-principles evolutionary crystal structure prediction. In addition to the well-known phases Ti$_2$N and TiN, we have uncovered three new intriguing structures at ambient conditions (Ti$_3$N$_2$, Ti$_4$N$_3$ and Ti$_6$N$_5$). At high pressures, two new phases \textit{Cmcm}-Ti$_2$N and \textit{I4/mcm}-TiN$_2$ were discovered. All those phases are mechanically and dynamically stable at ambient conditions. The calculated elastic constants \textit{C$_{ij}$} of Ti-N compounds are in good agreement with the available reported data. Other mechanical properties including bulk modulus, shear modulus, Young's modulus, and Poisson's ratio have been further computed from \textit{C$_{ij}$}. Among the studied Ti-N compounds, TiN$_2$ exhibits significant elastic anisotropy and possess the highest hardness of 27.2 GPa. We found a strong correlation between the mechanical properties and N content, e.g., the more N content, the more directional covalent Ti-N bonding and the less Ti-Ti metallic bonding, which lead to the enhancement of the hardness. The materials discovered here are attractive for technological applications because of a compromise between hardness and ductility, due to a peculiar interplay between metallicity and covalency.

\section*{Acknowledgements}
We thank the Natural Science Foundation of China (Grants No. 51372203 and No. 51332004), the Basic Research Foundation of NWPU (Grant No. JCY20130114), the Foreign Talents Introduction and Academic Exchange Program (Grant No. B08040), the National Science Foundation (Grants No. EAR-1114313 and No. DMR-1231586), DARPA (Grants No. W31P4Q1310005 and No. W31P4Q1210008), NSW (grant DMR-1231586) and the Government of the Russian Federation (Grant No. 14.A12.31.0003) for financial support. The authors also acknowledge the High Performance Computing Center of NWPU for the allocation of computing time on their machines.

\bibliographystyle{elsarticle-num}

\end{document}